\documentclass[twocolumn,amsmath,floatfix,amssymb,aps,prl,showpacs,letter, superscriptaddress]{revtex4-1}
\usepackage[pdftex]{graphicx}
\usepackage{graphicx} 

\usepackage{amssymb,amsmath}
\usepackage{dcolumn}
\usepackage{bm}
\usepackage{color,soul}
\usepackage{hyperref}

\begin {document}

\title{Spatially homogeneous ferromagnetism below the enhanced Curie temperature in EuO$_{1-x}$ thin films}

\author{Pedro M.~S.~Monteiro}
\email{pmdsm2@cam.ac.uk}
\affiliation{Cavendish Laboratory, Physics Department, University of Cambridge, Cambridge CB3 0HE, United Kingdom}

\author{Peter~J.~Baker}
\affiliation{ISIS Facility, STFC Rutherford Appleton Laboratory, Harwell Science and Innovation Campus, Oxon, OX11 0QX, United Kingdom}

\author{Adrian~Ionescu}
\affiliation{Cavendish Laboratory, Physics Department, University of Cambridge, Cambridge CB3 0HE, United Kingdom}

\author{Crispin~H.~W.~Barnes}
\affiliation{Cavendish Laboratory, Physics Department, University of Cambridge, Cambridge CB3 0HE, United Kingdom}

\author{Zaher~Salman}
\affiliation{Laboratory for Muon-Spin Spectroscopy, Paul Scherrer Institut, CH-5232 Villigen, Switzerland}

\author{Andreas~Suter}
\affiliation{Laboratory for Muon-Spin Spectroscopy, Paul Scherrer Institut, CH-5232 Villigen, Switzerland}

\author{Thomas~Prokscha}
\affiliation{Laboratory for Muon-Spin Spectroscopy, Paul Scherrer Institut, CH-5232 Villigen, Switzerland}

\author{Sean~Langridge}
\affiliation{ISIS Facility, STFC Rutherford Appleton Laboratory, Harwell Science and Innovation Campus, Oxon, OX11 0QX, United Kingdom}

\date{\today}


\begin{abstract}

We have used low energy implanted muons as a volume sensitive probe of the magnetic properties of EuO$_{1-x}$ thin films. We find that static and homogeneous magnetic order persists up to the elevated $T_{\rm C}$ in the doped samples and the muon signal displays the double dome feature also observed in the sample magnetization. Our results appear incompatible with either the magnetic phase separation or bound magnetic polaron descriptions previously suggested to explain the elevated $T_{\rm C}$, but are compatible with an RKKY-like interaction mediating magnetic interactions above 69~K.
\end{abstract}
\maketitle

The archetypal ferromagnetic semiconductor europium monoxide (EuO) is widely studied because it offers the ability to control spin-polarized electrical currents~\cite{Nagaev,Gregg,schmehl}. This possibility is limited by the low Curie temperature, $T_{\rm C} = 69$~K, of pristine EuO, but electronic doping by, for example, oxygen vacancies or the application of pressure can more than double the $T_{\rm C}$~\cite{Massenet1974,Barbagallo2010,Barbagallo2011,Tissen1987}. The physical origin of this elevated ordering temperature is controversial and has been discussed in terms of indirect exchange~\cite{Mauger1977,Burg2012}, bound magnetic polarons~\cite{Torrance,Liu2012} and chemical phase separation~\cite{Altendorf2011}. The existence of chemical phase impurity has been raised in bulk~\cite{Massenet1974} and thin film measurements~\cite{Altendorf2011}; this would seriously limit any technological applications~\cite{Wang2009,Miao2012}.

EuO crystallizes in the rocksalt Fm$\bar{3}$m crystal structure with a lattice constant of $5.144$~\AA~\cite{Eick1956}.
In bulk EuO there is a magnetic hard axis which is orientated along the [100] direction, whereas the easy axis is orientated along the [111] direction~\cite{Flosdorff1996} with a small magneto-crystalline anisotropy constant of $K_1 = \ -4.36 \times 10^{5}$~erg/cm$^{3}$ at 2~K \cite{PhysRev157448, PhysRev135A434}. For pristine EuO it is well established that the super-exchange interaction between the localised Eu \textit{4f} and O \textit{2p} states gives rise to a ferromagnetic order below its $T_{\rm C}$~\cite{Kasuya1970}.
The high degree of localization makes EuO a model Heisenberg ferromagnet, albeit with some momentum dependence of the exchange energy~\cite{Miyazaki2009}. The nearest and next-nearest neighbors exchange parameters were measured by neutron scattering as $J_1/k_B=0.606 \pm 0.008$~K and $J_2/k_B=0.119 \pm 0.015$~K~\cite{Passell1976}.

EuO has a magnetic moment of $7~\mu_{\rm B}$ per Eu atom~\cite{Barbagallo2010} and the half filled \textit{4f} level is separated from the conduction band by a band gap reported as 1.2~eV at 200~K~\cite{Schoenes} and 1.1 eV at 300~K~\cite{Gutherodt}. Its seven \textit{4f} electrons also create a Zeeman field which splits the conduction band about 0.6 eV~\cite{Steeneken} below its $T_{\rm C}$.
This leads to EuO being able to transport electrical currents with higher than $90\%$ spin polarization~\cite{schmehl}.
Doped EuO was also deposited epitaxially on Si (001), GaN (0001)~\cite{schmehl}, and on Ni (001)~\cite{EuONi} matching the carrier concentration of $\sim 10^{19}$~cm$^{-3}$ with Si, making it even more appealing for device applications~\cite{Wang2009,Muller2009,Haugen2008}.

Despite the extensive experimental and theoretical work on stoichiometric and Gd-doped EuO~\cite{AMauger1986}, far less attention has been paid to the effect of free carriers in oxygen deficient EuO$_{1-x}$, which act as an electron dopant and provide an additional magnetic interaction. Oxygen deficient EuO also undergoes a metal-to-insulator transition at 69~K~\cite{Torrance, Oliver1972,Penney,Altendorf2011} and exhibits a colossal magnetoresistance effect~\cite{Nagaev, Schmehl2011}.

One of the most plausible models~\cite{AMauger1986} for the elevated $T_{\rm C}$ proposes that in addition to the super-exchange, oxygen vacancies act as an electron donor ($n$-type doping) to the \textit{5d} band. Then the localized \textit{4f} electrons can couple via an indirect exchange mechanism mediated by the Eu \textit{5d} conduction electrons \cite{Kasuya1970}. This interaction is not the standard Ruderman-Kittel-Kasuya-Yosida (RKKY) interaction as observed in magnetic metals, but rather the semiconductor analogue and is a temperature dependent indirect interaction (RKKY-like) \cite{Mauger1977,Lee1984, Burg2012}. Another proposed model is based on bound magnetic polarons (BMP). This model was first proposed by Torrance {\em et al.}~\cite{Torrance} to explain the metal-to-insulator transition in oxygen deficient EuO. It assumes that oxygen vacancies are \textit{shallow} donors in which case one electron remains tightly bound, whereas the ``outer'' electron of the vacancy can be trapped by the vacancy under certain conditions such as temperature and carrier concentration \cite{Mauger1983}. Around 50~K and above $T_{\rm C}$ the trapped ``outer'' electron can cause the paramagnetic Eu$^{2+}$ spin to order. Thus the electron by localizing and ordering a small number of Eu$^{2+}$ spins gains magnetic free energy and forms a magnetic polaron, which being bound to an oxygen vacancy is termed BMP \cite{Nagaev, Liu2012}. However, it remained elusive whether the existence of the ``double-dome'' feature in the $M(T)$ data, $i.e.$ that the magnetization does not follow a single Brillouin function \cite{Massenet1974,Barbagallo2010,Altendorf2011,Liu2012}, originates either from two distinct phases with different $T_{\rm C}$'s or a single oxygen deficient EuO phase undergoing two magnetic transitions. In order to solve this ambiguity we have investigated the magnetic properties of the samples at a local level using low-energy muon spin relaxation and rotation (LE-$\mu$SR) measurements. Here spin-polarized positive muons are implanted inside the sample and the time evolution of their polarization is determined by measuring the direction of positrons emitted when the muons decay (for more details on LE-$\mu$SR see supplementary information).

The samples were deposited in a DC/RF magnetron sputtering system at room temperature. The deposition was performed by co-sputtering Eu$_2$O$_3$ and Eu. The oxygen deficiency was controlled by changing the relative deposition rate of Eu (see supplementary information).
In Fig.~\ref{SQUIDplots} we show the temperature dependent magnetization, obtained by SQUID magnetometry, for all three samples measured with an in-plane applied field of 10 kOe and the hysteresis loops at 5~K (inset). The stoichiometric EuO sample shows a paramagnetic response above $T_{\rm C}$, which was estimated by d$M(T)$/d$T$. The EuO$_{0.975}$ and EuO$_{0.91}$ samples have a $T_{\rm C}$ of 140~K and 136~K, respectively. For fields higher than 200~Oe, while $T_{\rm C}$ remains the same, a small paramagnetic response is present arising from the Pt capping layer.

Figure~\ref{SQUIDplots} shows a consistent reduction of the magnetic moment with temperature and a change in the coercive field: 54~Oe for the pristine sample, 79~Oe for the $2.5\%$ sample, and 236~Oe for the $9\%$ sample. The remanence is about 897~emu cm$^{-3}$ for the pristine and 842~emu cm$^{-3}$ for the $2.5\%$ sample, while it is about 300~emu cm$^{-3}$ for the $9\%$ sample at 5~K. Saturation of the pristine and the $2.5\%$ oxygen deficient films is obtained at applied fields of about 1000~Oe, while the $9\%$ sample can not be fully saturated even at 50~kOe (not shown). It was claimed in Ref.~\cite{Altendorf2011} that for lower applied fields the appearance of the ``double-dome" in the $M(T)$ measurement could remain unnoticed due to Eu clusters. However, we believe this is not an intrinsic feature of antiferromagnetic Eu clusters given that we observe hysteresis loops in the high temperature data of the oxygen deficient EuO samples (supplementary material) confirming ferromagnetic order above 69~K.

We previously reported \cite{Barbagallo2010} that the magnetic moment of EuO as estimated by self calibrating polarized neutron reflectivity is close to 7~$\mu_B$, higher than the ones obtained here by SQUID magnetometry, which has a much higher error on the determination of the magnetic moment. The error comes mainly from the determination of the area, thickness and roughness of the films.

\begin{figure}
 \includegraphics[width=9cm]{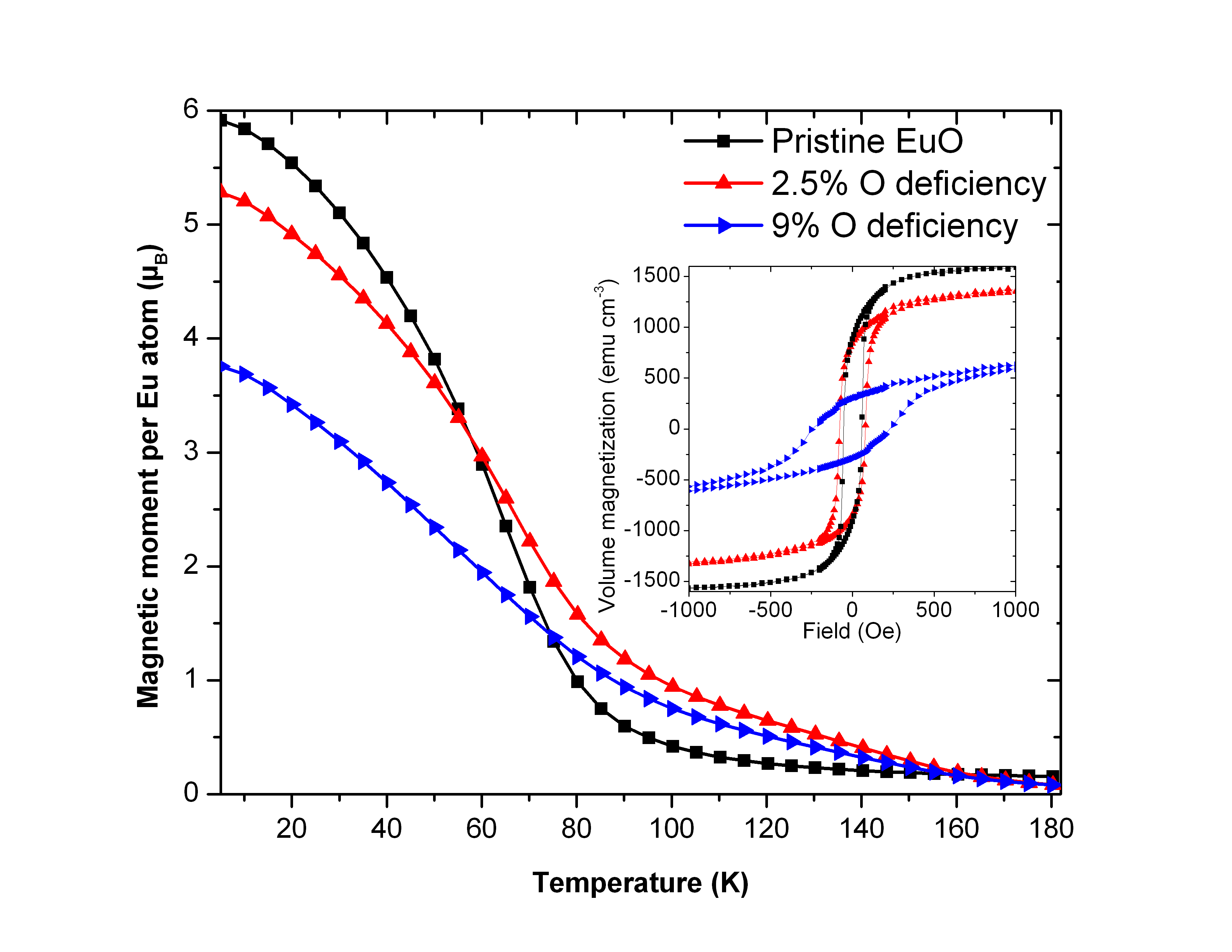}\\
\caption{The magnetic moment per Eu$^{2+}$ atom versus temperature at 10~kOe for the three samples and (inset) the volume magnetization as a function of the in-plane applied magnetic field at 5~K. The figure shows an increase of the coercive field and a decrease of the remanent magnetization as well as a reduction of the magnetic moment for increasing oxygen deficiency.}
\label{SQUIDplots}
\end{figure}

\begin{figure}
\includegraphics[width=9cm]{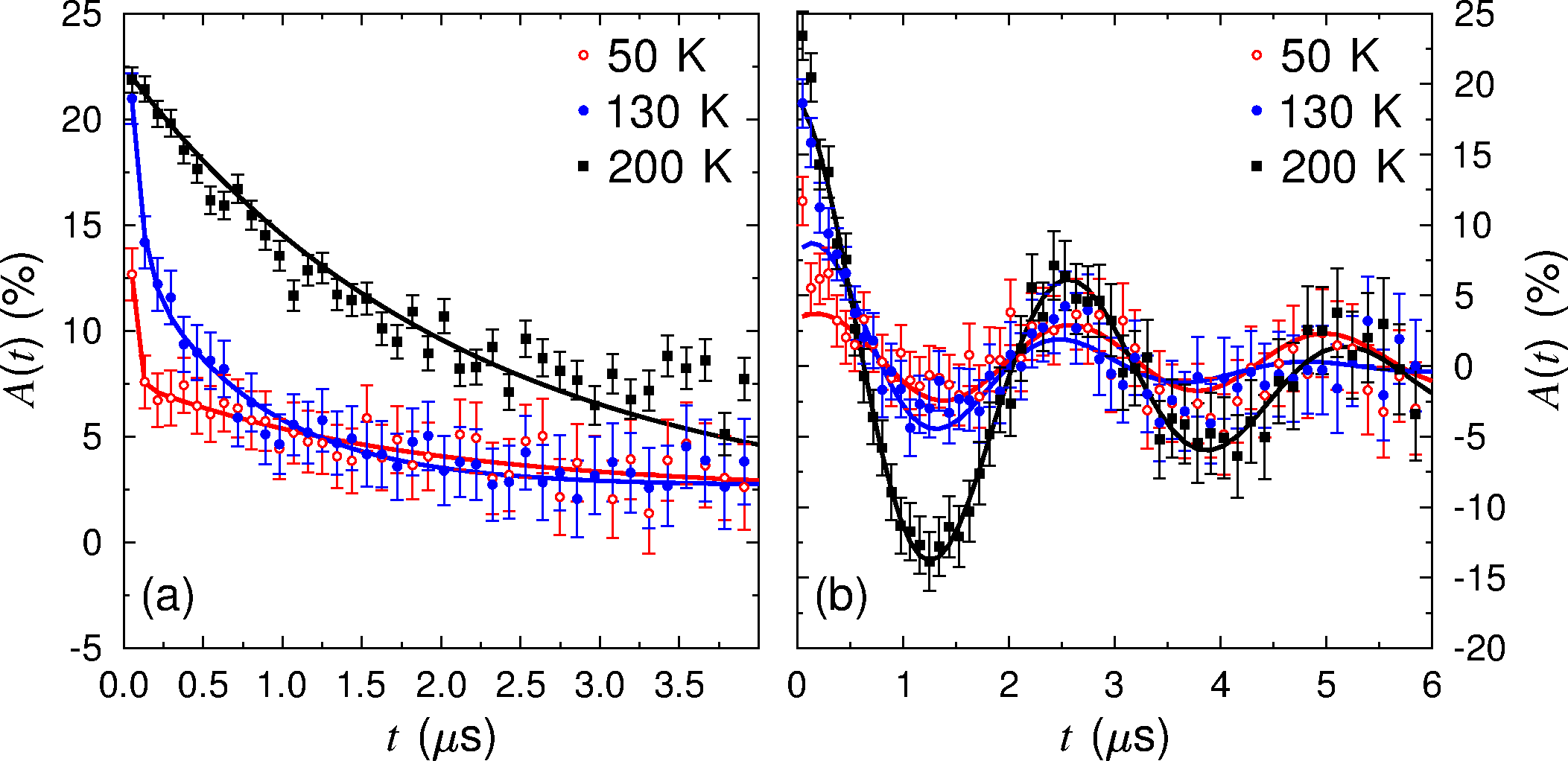}\\
\caption{Raw muon data for the EuO$_{0.91}$ sample:
(a) Zero-field measurements with the fits to Eq.~\ref{muZF}.
(b) Weak transverse field measurements with the fits to Eq.~\ref{muwTF}.
}
\label{mudatatrim}
\end{figure}

Our LE-$\mu$SR experiments comprised measurements in zero applied field (ZF) and a weak transverse field of 28.2\,G (wTF), where muons were implanted into the films at various energies between 6 and 14~keV (see supplementary information). The ZF measurement probes the static and dynamic properties of the spontaneous internal magnetic fields within the sample and the wTF enables the determination of its (strongly) magnetic volume fraction. For the ZF measurements we found that the raw data shown in Fig.~\ref{mudatatrim}~(a) were best described by the sum of two exponentially relaxing components below the $T_{\rm C}$ and a single exponential component above. In bulk EuO oscillations in the muon decay asymmetry are evident below $T_{\rm C}$~\cite{Blundell2010}, albeit with an apparent drop in the initial asymmetry compared to the paramagnetic state. We were not able to resolve oscillations consistently between different data sets for any of our samples. The lack of clear oscillations may be due to small variations in the internal fields caused by a distribution of grain boundaries and defects in the film.

We described the ZF data using the function:
\begin{equation}
A(t) = A_1 \exp(-\lambda t) + A_2 \exp(-\Lambda t) +A_{\rm bg},
\label{muZF}
\end{equation}
where the sum of $A_1$, $A_2$, and $A_{\rm bg}$ (background) is fixed by the geometry of the experiment and was fixed for each sample. The first term describes a slow relaxation of muon spins, which below $T_{\rm C}$ we can attribute to fluctuations of magnetic fields parallel to the muon spin polarization. The second term, describing the fast relaxation, is found to be zero above $T_{\rm C}$ and represents an incoherent precession of muons about fields perpendicular to their spin polarization. This relaxation rate $\Lambda$ will therefore vary with the size of those fields and gives us some insight into the magnetic order parameter, albeit far more limited than a well-defined oscillation frequency. We took $\lambda < \Lambda$ below $T_{\rm C}$ and set $A_2 = 0$, when a second relaxing component could not be resolved above $T_{\rm C}$. Below $T_{\rm C}$ we note that $A_2 / A_1 \sim 2$, which is the value anticipated for a polycrystalline ordered magnet. Using this we can estimate the fraction of the probed sample volume entering a magnetically ordered state at each temperature as $P_{\rm mag} = 1.5 \times A_2 / (A_1 +A_2)$.

The wTF measurements at low-temperature show a fast relaxation and a slowly-relaxing oscillation, which can be attributed to muons respectively experiencing the large spontaneous fields and the weak applied field. This gives us a more sensitive probe of the volume of the sample in which the muons are implanted that is magnetically ordered. To simplify the determination of this volume fraction we fitted the data omitting the first $0.25~\mu$s to the function:
\begin{equation}
A(t) = A_r \exp(-\eta t)\cos(\gamma_{\mu} B + \phi).
\label{muwTF}
\end{equation}
This describes the slow relaxing precession of muons not experiencing large internal magnetic fields within the magnetic volume of the sample. The relaxation rate $\eta$ describes how fast the precession is depolarized by a distribution of magnetic fields, $\gamma_{\mu}/2\pi = 135.5$~MHz/T is the muon's gyromagnetic ratio, $B$ is the magnetic field experienced by the implanted muons, and $\phi$ is a phase offset due to the spin rotation of the implanted muons relative to the detectors. Parameterizations accounting for the initial fast relaxation fitted over the whole data set give similar values for these parameters, but with a worse quality of fit, and small changes in the fitting time window lead to only minor differences in the parameters.

\begin{figure}
 \includegraphics[width=8cm]{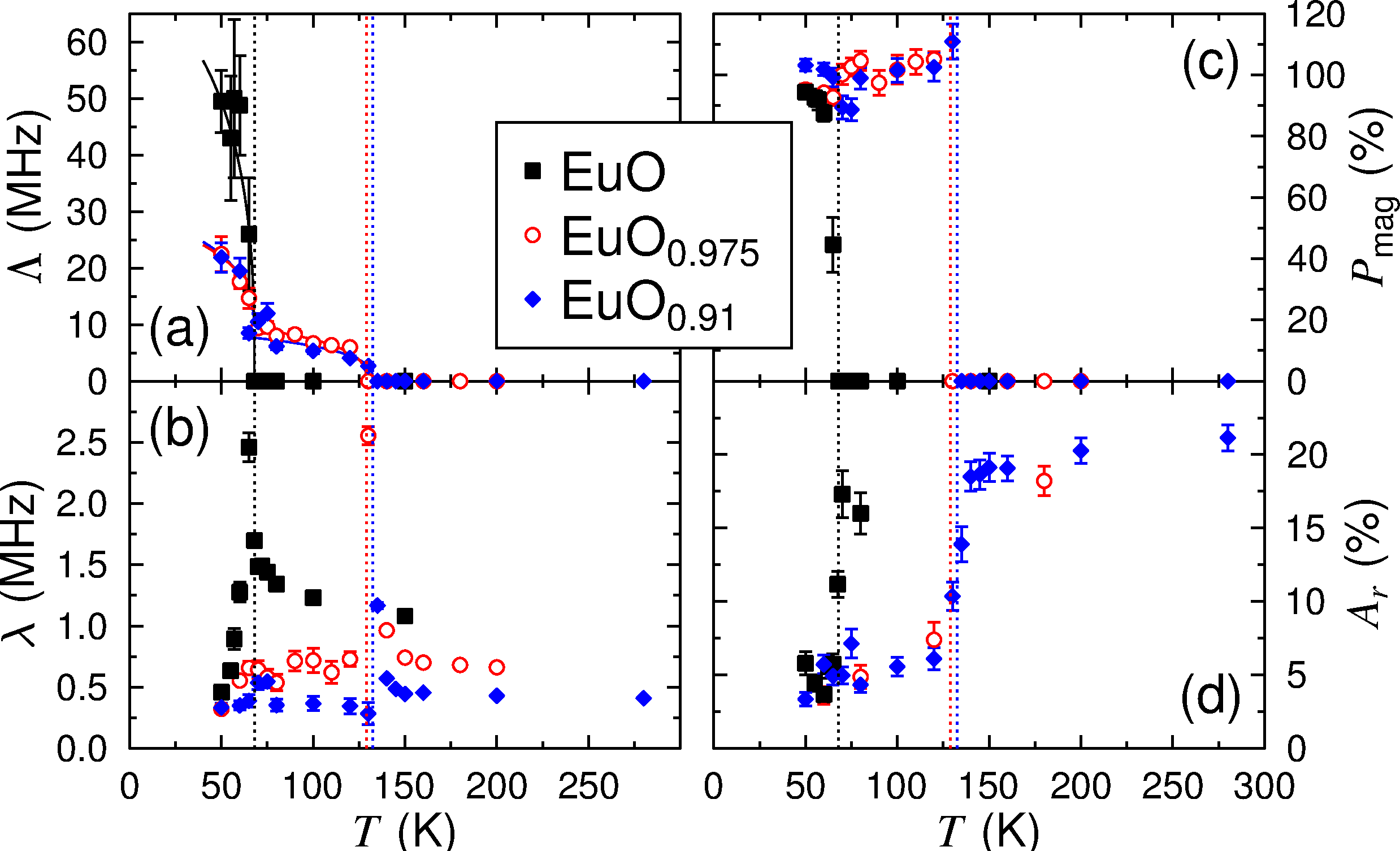}\\
\caption{Fitted parameters for the zero field muon data (a-c) and weak transverse field data (d):
(a) Fast relaxation rate, $\Lambda$. The lines are guides to the eye.
(b) Slow relaxation rate, $\lambda$.
(c) Static magnetic volume fraction, $P_{\rm mag}$, for the three samples.
(d) Relaxing asymmetry, $A_r$.
The vertical dotted lines denote the Curie temperatures for each sample.
}
\label{muparameters}
\end{figure}

The parameters resulting from fitting the ZF data with Eq.~\ref{muZF} are shown in Fig.~\ref{muparameters}~(a-c). In Fig.~\ref{muparameters}~(a) the faster relaxation rate $\Lambda$ is shown. This exhibits a similar temperature dependence to the magnetization recorded in our bulk measurements, wholly consistent with following a ferromagnetic order parameter with both techniques. Fig.~\ref{muparameters}~(b) shows the slower relaxation rate $\lambda$, which shows a similar peak at $T_{\rm C}$ to that recorded in the bulk measurements~\cite{Blundell2010} for the stoichiometric sample, and shows a peak at $T_{\rm C}$ in both of the oxygen-deficient samples. In Fig.~\ref{muparameters}~(c) $P_{\rm mag}$ values are shown as a function of temperature. These data show that quasistatic magnetic fields develop through close to the whole sample volume at $T_{\rm C}$, independent of the oxygen stoichiometry. These $T_{\rm C}$ values are slightly lower than those from our magnetization data. The uncertainty in the two measurements means the values are compatible with one another.

The wTF parameters are shown in Fig.~\ref{muparameters}~(d) and in Fig.~8 of the supplementary information. The value of $A_r$ [Fig.~\ref{muparameters}~(d)] drops on entering the magnetically ordered phase similarly in all of the samples, from around 20~\% to around 5~\%. This further demonstrates that the magnetic volume fraction is very similar in each of the three samples and ordering develops fully at the elevated Curie temperature in the doped samples. The residual low-temperature oscillating asymmetry can be attributed to muons implanted in the backing plate and in the non-magnetic layers of the sample, as all three samples had the same geometry, and the values are consistent with the $A_{\rm bg}$ values found when fitting the zero field data. $\eta$ peaks around $T_{\rm C}$ in each sample. There is a small but statistically significant growth in $B$ below $T_{\rm C}$ (Fig.~8 in the supplementary information). The sign of this change will be determined by the relative field directions of the contributions from the ordered moments and the sample magnetization, but it has the same sign and magnitude in all three samples showing that the magnetic ordering is similar, in accordance with the bulk measurements described above.

From both the ZF- and wTF-$\mu$SR results we obtain a static magnetic volume that is consistent 
with the full sample volume entering a magnetically ordered state at the Curie temperature obtained from both magnetization and $\mu$SR measurements. Together with the sharp transitions in all of our samples this appears to rule out the two chemical phase picture of EuO$_{1-x}$ with a significant distribution of elevated $T_{\rm C}$ values suggested in the earliest study~\cite{Massenet1974}.
We also note that the faster relaxation rate $\Lambda$, in following a similar temperature dependence to the bulk magnetization, suggests that the order parameter of the oxygen deficient samples grows faster below 69~K. The slow relaxation rate $\lambda$ also changes around 69~K but not in the same way as in the undoped sample. Since these changes cannot be associated with a change in the magnetic volume fraction of the sample another explanation must be sought.
One possibility would be entering a short range ordered state at the elevated $T_{\rm C}$ which coalesces into long range order below 69~K.
It is difficult to fully exclude this scenario since we do not observe coherent muon precession in either of the temperature regions. However, it is hard to reconcile such a picture with the sharp onset of static magnetism throughout each sample, accompanied by a distinct peak in $\lambda$ and $\eta$ at the elevated $T_{\rm C}$ in the doped samples, which is similar to that in the stoichiometric sample where the order may be expected to be solely long ranged.

The elevated $T_{\rm C}$ value in doped EuO is commonly discussed in terms of the BMP theory. This predicts magnetic phase separation as the number of BMPs should decrease progressively with temperature above the temperature at which long-range order disappears \cite{Mauger1983, Kimura2008}. Such a picture is not compatible with our results for the magnetically ordered volume fraction shown in Fig.~\ref{muparameters}~(c), where a complete and sharp change in the magnetic volume fraction is observed at the elevated $T_{\rm C}$ determined independently by the magnetization measurements. It has recently been argued by Liu $et$ $al.$ ~\cite{Liu2012} that features in the hysteresis, coercivity, and $M(T)$ data of their EuO$_{1-x}$ samples provided evidence for antiferromagnetically coupled BMP below $T_{\rm C}$.
Our data do not show these features. The coercivity does not exhibit a drop at 70~K (Fig.~4 of the supplementary information) and there is no drop in $M(T)$ at 20~K, (Fig.~\ref{SQUIDplots}) even though the dependence of $T_{\rm C}$ on oxygen deficiency found in Ref.~\cite{Barbagallo2010} implies that our samples should have a similar level of oxygen deficiency. Other recently published work~\cite{Barbagallo2010, Barbagallo2011, Altendorf2011} also show no sign of these features. This suggests that such effects are specific to the sample measured in Ref.~\cite{Liu2012} rather than generic properties of oxygen deficient EuO.

Detailed muon measurements~\cite{Brooks2004} of the prototypical magnetic polaron system EuB$_6$ provide a useful comparison with our EuO$_{1-x}$ data. In EuB$_6$ oscillations were clearly resolved below 12.6~K where neutron diffraction shows a significant spontaneous moment, but between that temperature and 15.5~K, where a small spontaneous moment is present, most of the oscillating asymmetry is replaced by a Gaussian term arising from static but random magnetic fields. Above 9~K there is a growth in the relaxation rate $\lambda$ consistent with two-magnon processes and no divergence in the exponential relaxation rate was evident at either of the magnetic transitions. Two exponential components are evident up to above 100~K in data with a better resolution to slow relaxation rates~\cite{Brooks2004}. In our EuO$_{1-x}$ data we are not able to resolve oscillations. However, there is no evidence for a three-component relaxation between 70~K and $T_{\rm C}$ that would be comparable with the data recorded in the intermediate temperature range in EuB$_6$ or two-components in the relaxation above $T_{\rm C}$. Such behavior above $T_{\rm C}$ could be outside our time resolution but comparing the amplitudes observed in the undoped and doped samples shows no significant differences, which is more likely to be consistent with a single relaxing component and a constant background. The exponential relaxation rate $\lambda$ in EuO$_{1-x}$ also shows a clear peak at $T_{\rm C}$ in each sample as in the stoichiometric sample. This is usually associated with the development of long-range magnetic order, but is not observed in EuB$_6$, although the critical behavior of Heisenberg ferromagnets can inhibit such a peak. Thus our doped samples appear to show the same phenomena around their elevated $T_{\rm C}$ values as the undoped sample does around 69~K, without further phenomena that can be associated with polarons. Muon-induced polarons (bound electron around a positive muon) have also been suggested to occur in EuO on the basis of an anomalously large relaxation rate at high temperatures~\cite{Storchak2010}. While we observe broadly comparable values for the muon spin relaxation rate, they are compatible with previous bulk measurements and theoretical calculations which do not require polarons, which were compared in Ref.~\cite{Blundell2010}.

An alternative picture for the increased $T_{\rm C}$ in EuO$_{1-x}$ has been provided by Mauger~\cite{Mauger1977,Mauger1983}. This argues that for high doping concentrations the excess electrons populate the conduction band and give rise to an RKKY-like interaction that is responsible for the increase in $T_{\rm C}$.

The electronic phase diagram of EuO$_{1-x}$ was calculated by comparing the Mott delocalization criterium with this model~\cite{Mauger1983}.

This predicts that only in a small region of doping concentrations around $2\times 10^{20}$ electrons/cm$^3$ will EuO present a mixed state between metallic and insulator phases above 70~K. This means that only the insulating and partially mixed region will contain BMPs. In the metallic phase the magnetic polarons are ionized and electrons are delocalized in the conduction band. Assuming that each vacancy in our films provides only one ``outer" electron, $e.g.$ the $2.5$~\% sample has a doping concentration of around $7.3\times 10^{20}$ electrons/cm$^3$, well above the mixed state where BMPs would be active.
All these arguments support the idea that the mechanism responsible for the increase in the $T_{\rm C}$ in our films is the RKKY-like interaction and not BMP.

In this letter we demonstrate that stoichiometric and oxygen-deficient thin films of EuO develop static magnetic order below $T_{\rm C}$ throughout their volume. The behavior of pristine and doped samples around $T_{\rm C}$ is similar even though the ordering temperature is almost doubled in both oxygen-deficient films. We find no evidence of chemical phase separation. The double-dome feature widely observed in the magnetization of oxygen-deficient EuO$_{1-x}$ is also evident in our $\mu$SR data. From these results we conclude that the change in the magnetic behavior around 69~K in the oxygen-deficient films is due to a crossover between the dominance of the direct exchange present in stoichiometric EuO and the RKKY-like interaction due to electrons delocalized in the $5d$ conduction band that increases $T_{\rm C}$.

\begin{acknowledgments}

We thank Francis Pratt, Stephen Blundell, and Eugene Kogan for useful discussions. The authors acknowledge the financial support from the STFC. P.M.S.M. thanks the EPSRC, U.K., and Funda\c c\~{a}o para a Ci\^{e}ncia e a Tecnologia (SFRH/BD/71756/2010), Portugal. Part of this work was performed using the LEM instrument~\cite{pt1,pt2} at the Swiss Muon Source, Paul Scherrer Institute, Villigen, Switzerland.
\end{acknowledgments}

\bibliography{EuOx}

\end{document}


\maketitle

\section{Experimental methods}
\subsection{Sample preparation}
The polycrystalline samples were deposited at room temperature using a CEVP magnetron sputtering system with a base pressure of 5$\times$10$^{-9}$ Torr. Co-deposition was performed using two targets: a 99.99$\%$ pure Eu$_{2}$O$_{3}$ and a 99.99$\%$ pure Eu target. The EuO$_{1-x}$ films were co-deposited while maintaining the RF power constant at 75 W for the Eu$_{2}$O$_{3}$ target and changing the DC deposition current for the Eu target between 0.05 and 0.15 A. The growth was performed in an Ar$^{+}$ plasma at a pressure of 2 mTorr with a flow rate of 14 sccm. The substrates used were 1'' Si (001) with a native oxide layer. A Pt layer was deposited between the substrate and the EuO$_{1-x}$ film and another on the top as a capping layer to prevent further oxidation of the film. The nominal sample structure was Si(001)/SiO$_2$(1.4 nm)/Pt(5 nm)/EuO(100 nm)/Pt(5 nm).


\subsection{Bulk magnetic measurements}

The SQUID magnetometry measurements were performed in a Quantum Design MPMS 2 system. Each sample was cooled in a field of 50 Oe to 5 K followed by the $M(H)$ measurements done at increasing temperatures. Finally each sample was cooled down at zero field and the temperature dependent measurements where performed at different fields.

\begin{figure}
 \includegraphics[width=12cm]{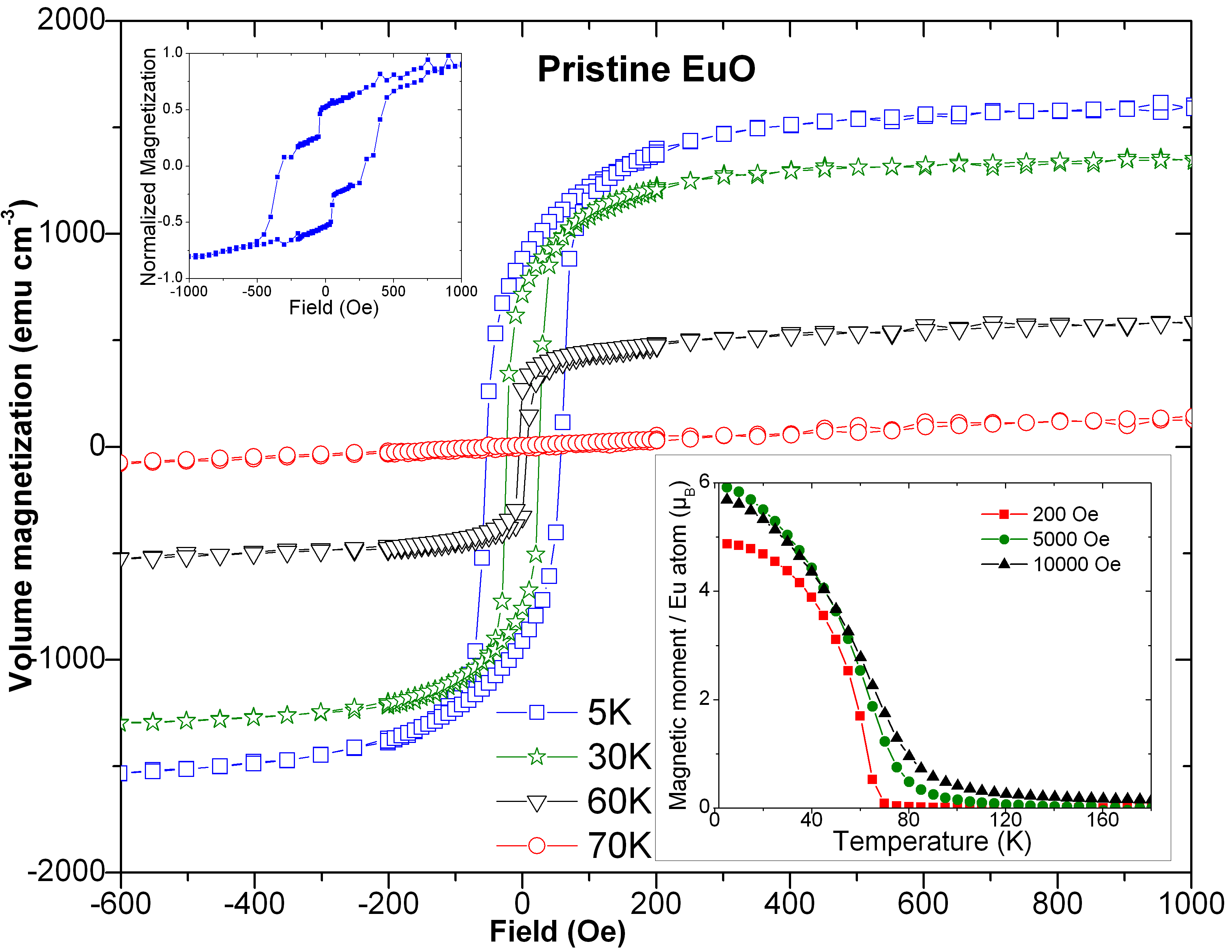}\\
\caption{Volume magnetization as a function of the in-plane applied magnetic field at different temperatures and (bottom inset) the magnetic moment per Eu$^{2+}$ atom versus temperature at 10 kOe, 5 kOe and 200 Oe for the pristine sample. The upper inset shows the out-of-plane normalized magnetization versus field at 5 K.}
\label{SQUID0}
\end{figure}

\begin{figure}
 \includegraphics[width=12cm]{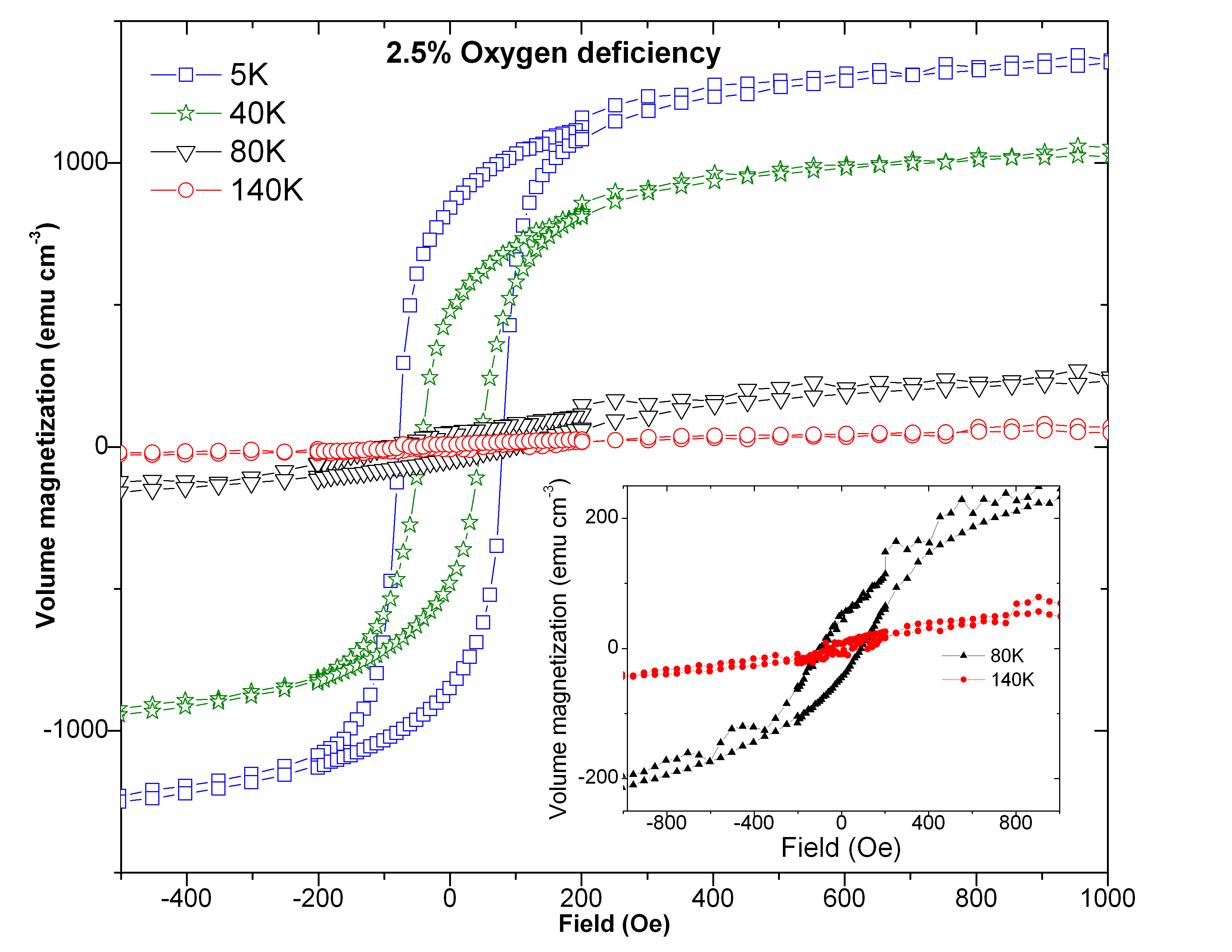}\\
\caption{Volume magnetization as a function of the in-plane applied magnetic field at different temperatures and (inset) an enlarged view of the higher temperature data showing a finite coercivity and remanence at 80 K and an almost paramagnetic response at 140 K for the $2.5\%$ oxygen deficient sample.}
\label{SQUID2.5}
\end{figure}

\begin{figure}
 \includegraphics[width=12cm]{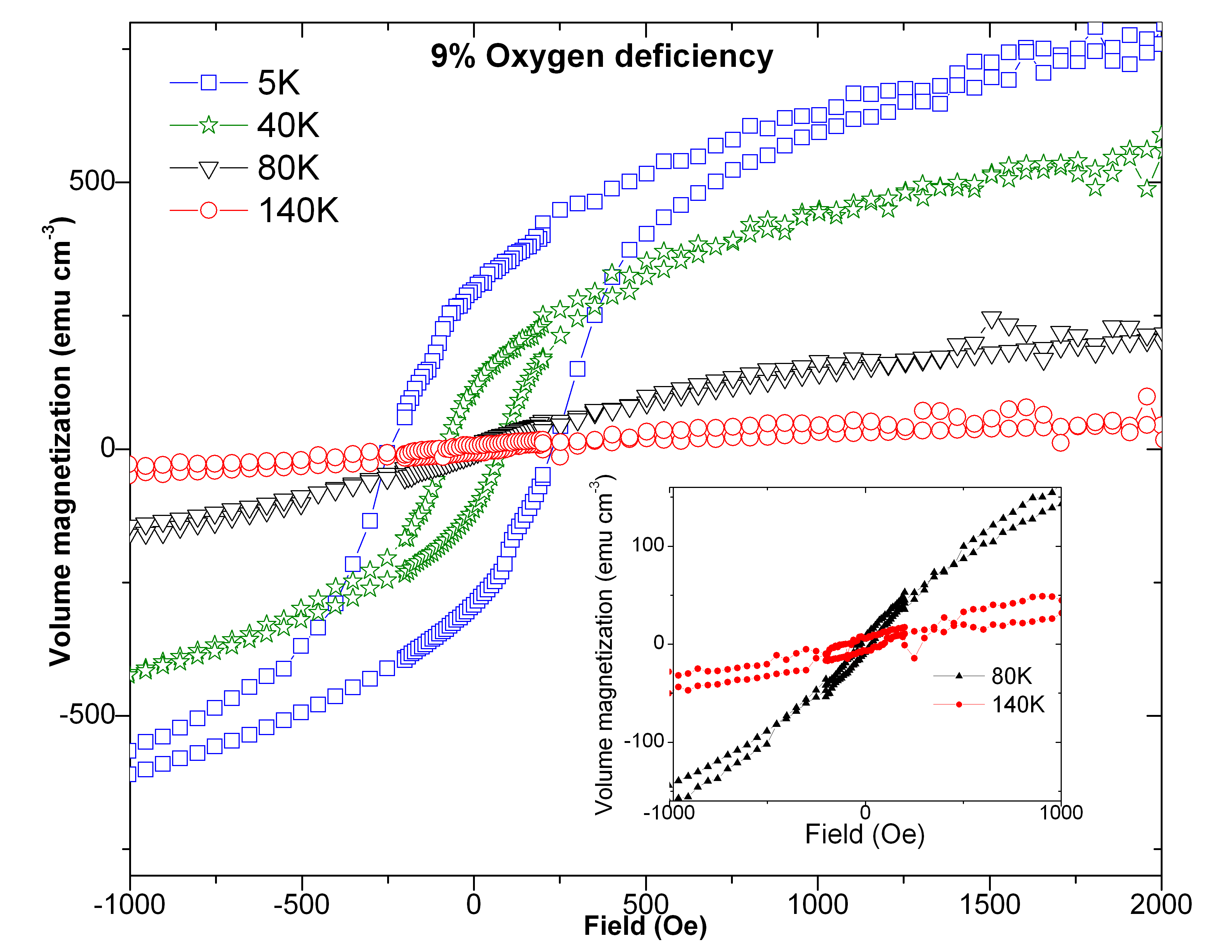}\\
\caption{Volume magnetization as a function of the in-plane applied magnetic field at different temperatures and (inset) an enlarged view of the higher temperature data showing a finite coercivity and remanence at 80 K and a paramagnetic response at 140 K for the $9\%$ oxygen deficient sample.}
\label{SQUID9}
\end{figure}

Figure~\ref{SQUID0} to \ref{SQUID9} show the hysteresis loops $M(H)$ at different temperatures for the pristine, $2.5\%$ and $9\%$ oxygen deficient samples. The inset to Fig. \ref{SQUID0} represents the $M(T)$ data for different applied fields. The pristine sample exhibits a paramagnetic response above 69 K. The inset to Fig. \ref{SQUID0} shows that for fields below 200 Oe the $T_{\rm C}$ remains the same however it did not reach the saturation magnetization. Above 200 Oe there is some paramagnetic signal at higher temperatures most probably from the $Pt$ layers. The upper inset shows the out-of-plane normalized magnetization versus field at 5 K. It exhibits a harder magnetic axis than the in-plane data with larger coercive fields and competing anisotropy contributions. The XRD data showed that the films are heavily (100) oriented (textured) in the out-of-plane direction which represents the hard magnetic axis in bulk EuO.

The $2.5\%$ (Figure \ref{SQUID2.5}) and $9\%$ (Figure \ref{SQUID9}) samples have a $T_{\rm C}$ as high as 140 K and 136 K, respectively. The insets show an enlarged view of the high temperature $M$($H$) data, where a finite coercivity and remanence at 80 K is present in both oxygen deficient samples underlining the existence of long range ferromagnetic order.

Figure 4 shows the temperature dependence of the coercivity of the $2.5\%$ and $9\%$ oxygen deficient samples. The $2.5\%$ shows a decrease in the coercivity between 5 K and 60 K followed by an increase at higher temperatures. The $9\%$ data shows a monotonic decrease of the coercivity with temperature. We attribute this changes to the variations of the in-plane crystal anisotropies.

\begin{figure}
 \includegraphics[width=12cm]{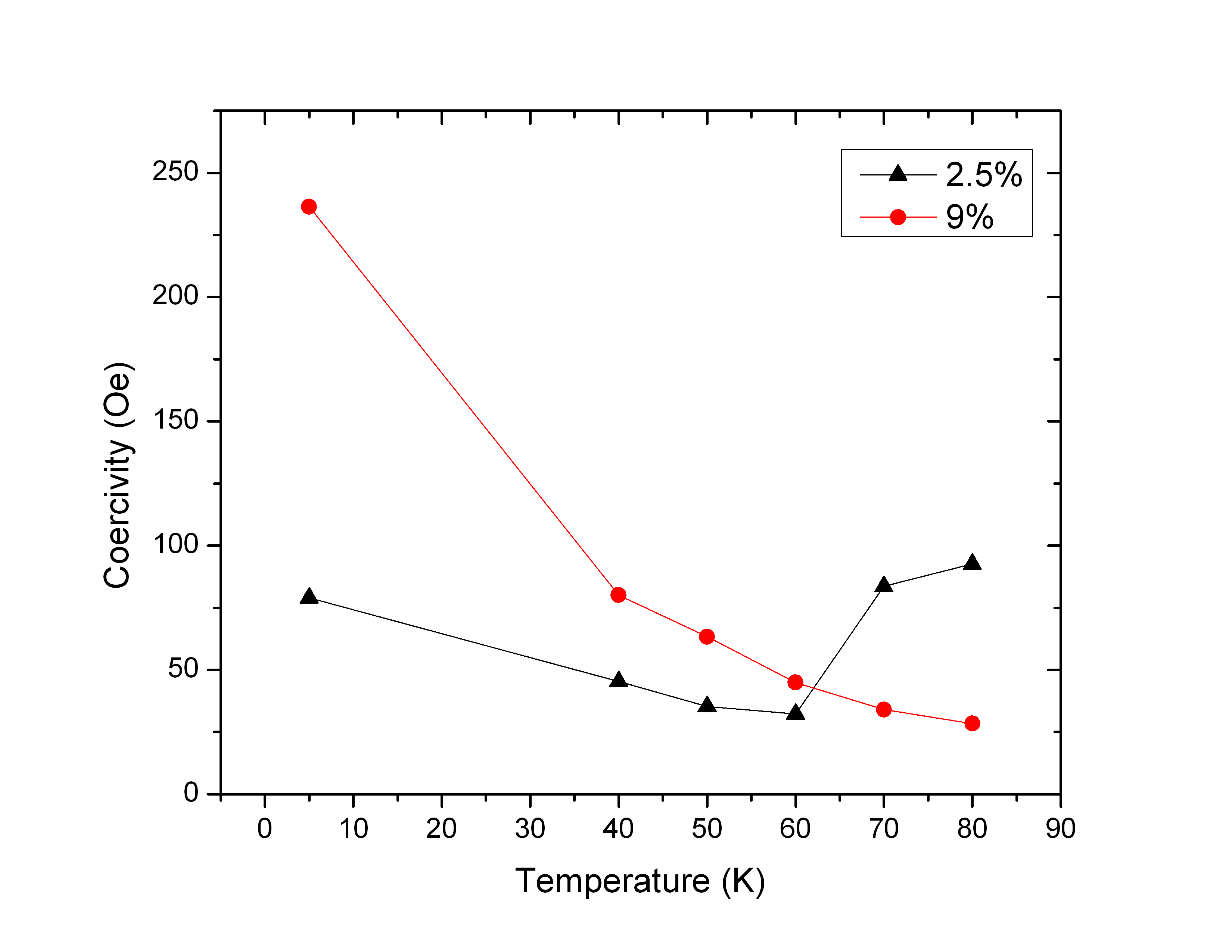}\\
\caption{Coercivity of EuO$_{1-x}$ films has a function of temperature.}
\label{SQUIDcoer}
\end{figure}

\subsection{X-ray reflectivity (XRR)}

\begin{figure}
 \includegraphics[width=12cm]{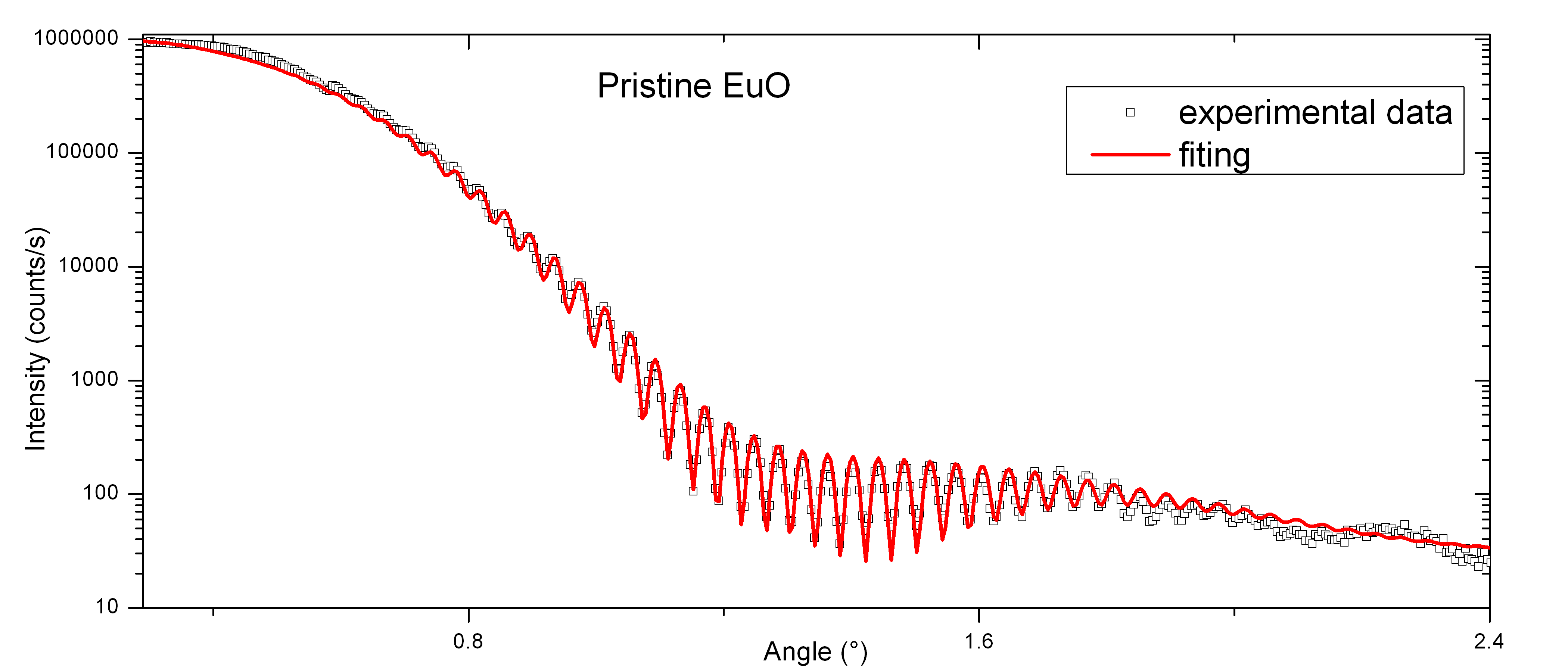}\\
\caption{X-ray reflectivity data (open symbols) and fitting (straight line) of the pristine sample.}
\label{xrr0}
\end{figure}

\begin{figure}
 \includegraphics[width=12cm]{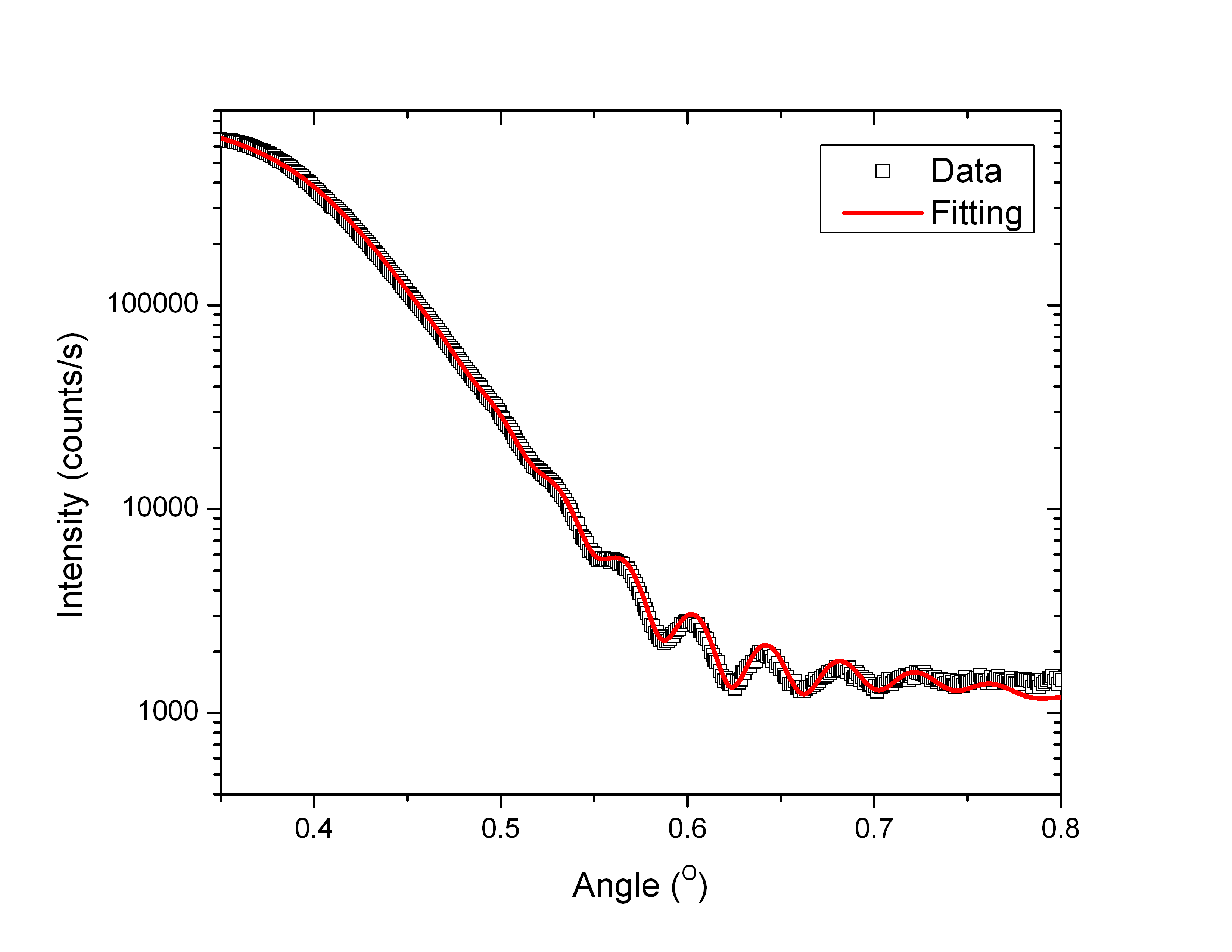}\\
\caption{X-ray reflectivity data (open symbols) and fitting (straight line) of the $2.5\%$ oxygen deficiency sample.}
\label{xrr2.5}
\end{figure}

The XRR data for the pristine sample was acquisitioned on a Panalytical
\newline
PW3050/65 X'Pert PRO HR horizontal diffractometer in low resolution mode with a step size of 0.001 degrees. The XRR data for the 2.5$\%$ oxygen deficient sample was measured on a Bruker D8 HRXRD in low resolution mode with a step size of 0.001 degrees. No useful XRR data has been obtained on the 9$\%$ oxygen deficient sample due to a large interface roughness. Tables 1 and 2 show the fitting parameters obtained from the two reflectivity curves presented in figures \ref{xrr0} and \ref{xrr2.5}, respectively. An increase in the interface roughness and structural disorder (densities are altered from their bulk values) for the non stoichiometric sample is apparent due to the quicker deposition rate of Eu. This trend is continued for the 9$\%$ oxygen deficiency sample, which showed no useful reflectivity data. While the nominal thickness of all the EuO films was 100 nm, we obtained 97.9 nm for the stoichiometric film and 89.3 nm for the 2.5$\%$ sample.

\begin{table}[htbp]

\center
\begin{tabular}{cccc}
\hline
\ \ Layer\ \ & \ \ Density (g/cm$^3$) \ \ &\ \ Thickness (nm) \ \  & \ \ Roughness (nm)\ \ \\
\hline \hline
 $Pt$  & 21.4   & 4.2     & 1.0 \\
                          $EuO$ & 8.2 & 97.9        & 1.2 \\
 $Pt$  & 21.4    & 3.7      & 0.7\\
                                $Si$   & 2.64       & $\infty$  & 0.2 \\
\end{tabular}
\label{tab:xrr0}
\caption{X-ray reflectivity fitting parameters for pristine EuO. The fit to the data shows a strong agreement with the experimental results.}
\end{table}

\begin{table}[htbp]

\center
\begin{tabular}{cccc}
\hline
\ \ Layer\ \ & \ \ Density (g/cm$^3$) \ \ &\ \ Thickness (nm) \ \  & \ \ Roughness (nm)\ \ \\
\hline \hline
 $Pt$  & 18.7   & 0.6     & 3.3 \\
 $EuO$ & 8.5 & 89.3        & 3.2 \\
 $Pt$   & 17.6  & 4.6      & 0.9\\
 $Si$    &   2.6   & $\infty$  & 0.8 \\
\end{tabular}
\label{tab:xrr2.5}
\caption{X-ray reflectivity fitting parameters for 2.5$\%$ oxygen deficiency of EuO. The fit to the data shows a strong agreement with the experimental results.}
\end{table}

\subsection{Muon spin relaxation measurements}

Muon spin relaxation and rotation ($\mu$SR) measurements were carried out using the low-energy muon spectrometer (LEM) at the Paul Scherrer Institute, CH. In $\mu$SR experiments \cite{Blundell2010} spin polarized positive muons are implanted into the sample, where they stop rapidly at interstitial sites of high electron density, and their spin direction evolves in the magnetic field at their stopping site. Each implanted muon decays with a lifetime of 2.2 $\mu$s emitting a positron preferentially in the direction of its spin at the time of decay. The evolution of their spin polarization as a function of time is determined by measuring the direction of the emitted positrons. In LEM experiments, accelerating electric fields are used to control the implantation depth of the muons after they have passed through a cryogenic moderator.

\begin{figure}
 \includegraphics[width=12cm]{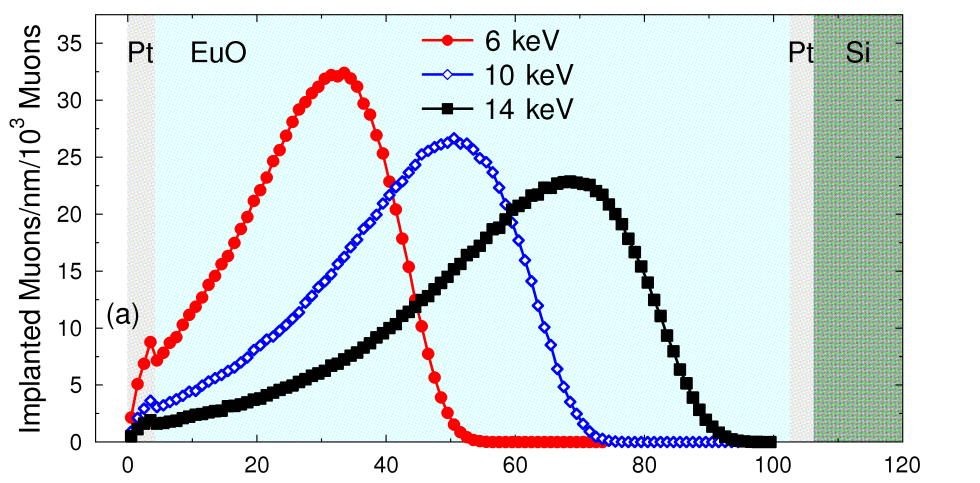}\\
\caption{Muon implantation depth calculations for the EuO sample at the three implantation energies used during our measurements.}
\label{mu}
\end{figure}

To determine where the muons would implant within the samples we carried out Monte Carlo simulations using the TRIM:SP program \cite{eckstein,pt3}. Results of these calculations for the energies used in our measurements are shown in Fig.~\ref{mu} for the EuO sample. The other samples give similar profiles. Our measurements at 6, 10, and 14~keV showed small changes consistent with the calculated stopping profiles. To maximize the proportion of muons stopping within the EuO$_{1-x}$ layer of the samples we used an implantation energy of 14~keV for the temperature dependent measurements. At $14$~keV approximately $1$\% of the muons stop in the front Pt capping layer, $3$\% are reflected, and the other $96$\% stop within the EuO$_{1-x}$ layer.

\begin{figure}
 \includegraphics[width=12cm]{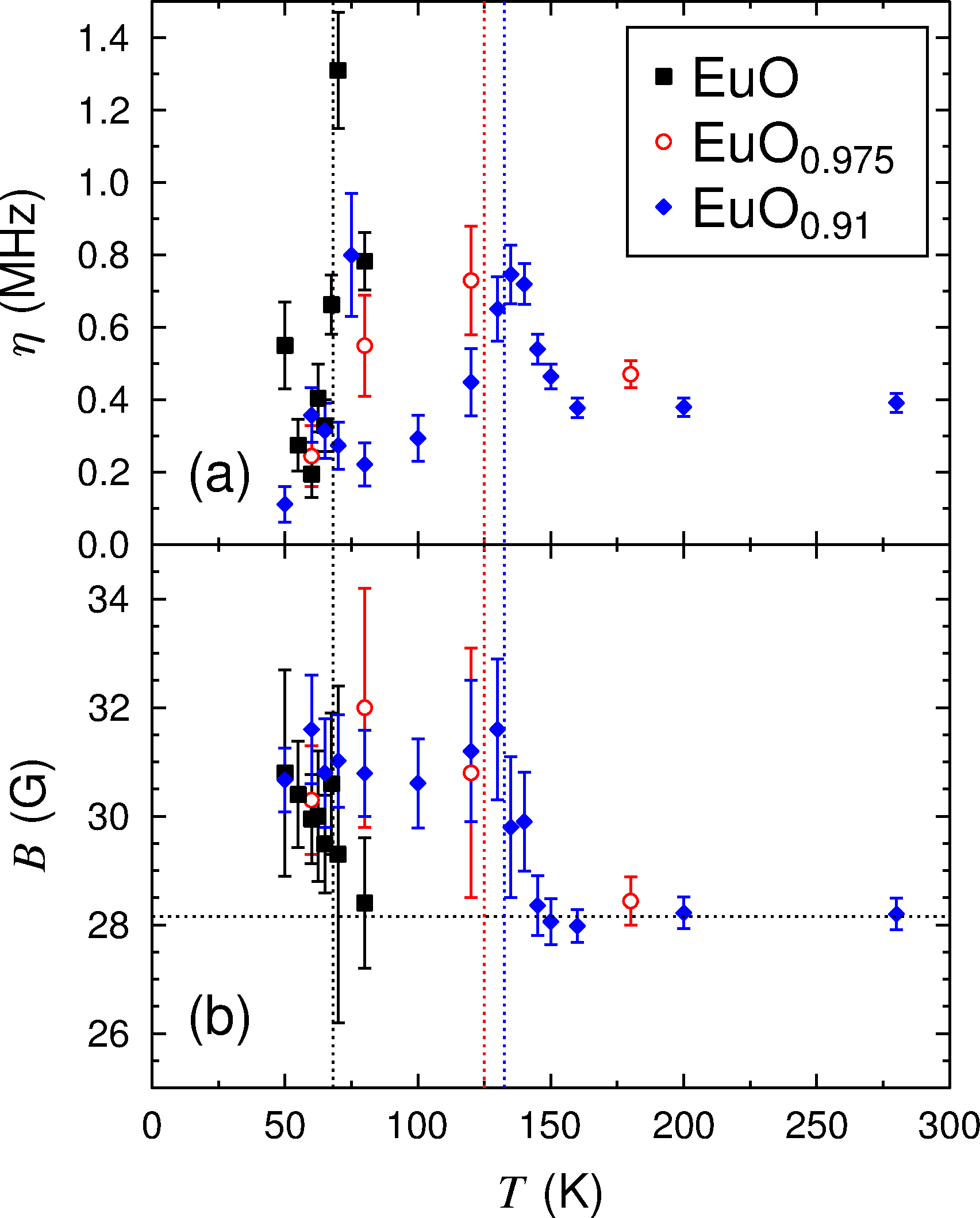}\\
\caption{Parameters derived from fitting Eq.~2 to the weak transverse field data.
(a) Relaxation rate $\eta$ and
(b) Magnetic field $B$ experienced by the muons.
}
\label{supp_params}
\end{figure}

Fig.~3~(d) in the main text shows the relaxing asymmetry for a weak transverse field of 28.2~G. In Fig.~\ref{supp_params} we show the other two parameters described in the main text, $\eta$ and $B$, the relaxation rate and magnetic field experienced by the muons. $\eta$ shows a peak at $T_{\rm C}$ in each of the samples and a more subtle change around $70$~K in the two oxygen deficient samples. The magnetic field experienced by the muons increased upon cooling below $T_{\rm C}$ in each sample and does not change significantly around $70$~K in the oxygen deficient samples.

\bibliography{EuOx}